\documentclass[conference]{IEEEtran}
\usepackage{cite}
\usepackage{amsmath,amssymb,amsfonts}
\usepackage{algorithmic}
\usepackage{graphicx}
\usepackage{textcomp}
\usepackage{xcolor}
\usepackage[ruled]{algorithm2e}
\def\BibTeX{{\rm B\kern-.05em{\sc i\kern-.025em b}\kern-.08em
    T\kern-.1667em\lower.7ex\hbox{E}\kern-.125emX}}
\begin{document}

\title{The Effect of Data Marshalling on Computation Offloading Decisions}

\author{\IEEEauthorblockN{Julio A. Reyes-Munoz}
\IEEEauthorblockA{\textit{Electrical and Computer Engineering} \\
\textit{The University of Texas at El Paso}\\
El Paso, TX, USA \\
jareyesmuno@miners.utep.edu}
\and
\IEEEauthorblockN{Michael P. McGarry}
\IEEEauthorblockA{\textit{Electrical and Computer Engineering} \\
\textit{The University of Texas at El Paso}\\
El Paso, TX,  USA \\
mpmcgarry@utep.edu}
}

\maketitle

\begin{abstract}
We conducted an extensive set of experiments with an offloading testbed to understand the impact that data marshalling
techniques have on computation offloading decisions. We find that the popular JSON format to marshall data between client
and server comes at a significant computational expense compared to a minimalistic raw data transfer. The computational time
is significant in that it affects computation offloading decisions in a variety of conditions. We outline some of these 
conditions.
\end{abstract}


\section{Introduction}
The idea of providing computing-as-a-utility is closely related with the origins of time-sharing operating systems at MIT. 
During a lecture in 1961 about the trend of time-sharing computers, i.e. computers that serve many users through console 
terminals, John McCarthy~\cite{M1962} coined the concept of computing as a utility. He envisaged future computers organized to 
provide their services to subscribers connected through telephone lines on a pay-per-use basis. This concept inspired Fernando 
Corbato to develop the Compatible Time Sharing System (CTSS) operating system to make a computer available to several 
users at the same time~\cite{CMD1962}. Some years later, in 1977, Madnick described the hardware and software requirements for 
an information utility system, where specialized computers provide storage and processing power to personal computers~\cite{M1977}. 

There are several recent developments that have made computing-as-a-utility a viable business and thereby intensifying research interest 
in this area. The first is the ubiquity of high bandwidth at the edge of the Internet that allows the movement of large amounts of data
within a reasonable time frame. The second is the advancement of virtualization technologies that permit rapid provisioning of computing
and storage resources with flexible granularities. The third is the advancement of data center technologies that leverage economies of
scale that significantly lower the cost per unit of computing or storage. We urge readers to consult \cite{AFGJKKLPRSZ2010} for a detailed discussion
of these and other developments that have made computing-as-a-utility viable. With computing-as-a-utility becoming commonplace there is
great interest in investigating ways to utilize it. We feel that one of the most compelling uses for computing-as-a-utility is for
computation offloading.

Computation offloading allows resource constrained computers, such as smartphones and tablets, to use computing-as-a-utility for the 
remote execution of resource intensive computing tasks. By using more powerful computers available at data centers or in 
cloudlets~\cite{SBCD2009} at the edge of the network, task completion times can be reduced or power consumption can be shifted. 
Computation offloading like computing-as-a-utility has a rather long history as a concept having been discussed since 1998, when job 
migrations for execution at static computing devices were proposed with the objective of reducing power consumption on mobile devices 
\cite{RRPK1998, OH1998}. However, with its viability there is significant recent interest in computation offloading systems and the types 
of computation tasks that would most benefit from computation offloading. To reduce completion time, the sum of the data transmission 
between the local (e.g., mobile device) and remote hosts (e.g., data center) along with the remote execution time must be less than the 
local execution time. A \textbf{client/server} application fixes the decision to offload computing (i.e., to the server); in some scenarios
this provides the best solution. In other scenarios conditions are dynamic and a dynamic computation offloading decision will provide
better performance. In this manuscript, we study systems that support a dynamic computation offloading decision.

\begin{figure}[htbp]
	\centering
	\includegraphics[scale=0.3]{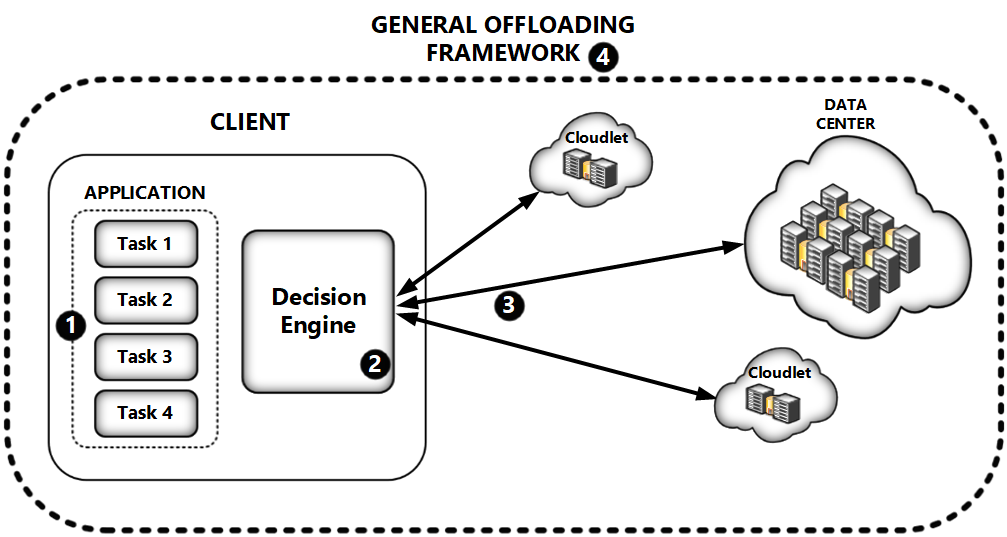}
	\caption[Four offloading decision questions represented on a general framework.]{{\scriptsize The four computation offloading questions are illustrated in a general architecture for offloading. In 1, the application is partitioned into several tasks, some of which are going to be offloaded, corresponding to the \textbf{what} question. In 2, the offloading decision is taken depending on several environmental and system variables, in order to decide \textbf{when} to offload subject to a set of objectives. In 3, the compute intensive task is executed at remote resources, which could be small servers close to the user, or even in data centers remotely located. Hence, \textbf{where} to execute should be decided. Finally, in 4 is shown the general offloading framework, which is actually \textbf{how} computation offloading is implemented, including the system architecture, the programming environments, and the techniques used for handling data and remote communication.}}
	\label{fig:questions}
\end{figure}

We view a computation offloading system as one that addresses four questions, illustrated in Fig. \ref{fig:questions}:

\begin{itemize}
	\item{What should be offloaded?}
    \item{When is offloading beneficial?}
    \item{Where is the appropriate offloading location?}
    \item{How to offload?}
\end{itemize}

\subsection{Related work}
Several works on computation offloading explored these four questions. Several seminal frameworks offload tasks encapsulated in functions, like 
in the case of Spectra~\cite{FPS2002} and Chroma~\cite{BSPO2003}, or VM replicas like Slingshot~\cite{SF2005}, when the user looks to reduce 
execution time. These frameworks use heuristic solvers combined with resource monitoring. Recent work proposes offloading to shift energy 
consumption to the cloud, like MAUI~\cite{CBCWSCB2010}, or combining both objectives in the case of CloneCloud~\cite{CIMNP2011}, 
Cuckoo~\cite{KPKB2012}, and Thinkair~\cite{KAHMZ2012}. These frameworks offload methods too, except for CloneCloud that keeps a VM updated 
at the cloud where a mobile device clone is running. In many cases, the offloading location is a remote data center, like in the case of MAUI, 
CloneCloud, Thinkair, and Cuckoo, or to cloudlets, like Spectra, Chroma, or Slingshot. Finally, some frameworks 
manage code synchronization besides the actual transmission of the data, like Thinkair and Cuckoo, whereas the other frameworks only 
transfer the I/O data required by the computing task. Data transmission implies some overhead for preparing the data. For example, MAUI uses 
XML for marshalling, and the Cloudlets system~\cite{SLMHS2012} sends VM overlays through HTTP. Most of the authors are aware of the marshalling 
overhead, but they do not measure the extent of this variable, nor do they talk about the specific implementation or technique used for 
marshalling.

Melendez, et. al. \cite{MM2017}, developed an inequality for offloading decisions that relates the parameters of an offloading system with 
the arithmetic intensity, i.e. the number of instructions executed over the amount of data moved through the network, of a computational job. 
This expression can be used to ascertain which categories of computation benefit from offloading in a particular offloading system, or in a 
similar manner, the offloading system requirements for a certain category of computations to benefit from offloading. This inequality
does not consider the overhead of formatting data for marshalling.

The main contribution of this work is a quantitative analysis of how the data handling and transmission techniques, i.e. data marshalling, 
of a particular offloading system affect the offloading decision, regarding whether offloading is convenient or not for reducing completion 
time. A set of experiments were run over a physical offloading testbed in order to uncover hidden low level factors that affect the 
offloading decision. It was discovered that the data marshalling/un-marshalling has a significant impact, especially when the solution lies 
close to the border between local and remote execution. Matrix operations with different input sizes were used to run the experiments, and 
the completion times of these jobs were compared by using two different methods of data serialization: JSON vs RAW data.

\subsection{Outline}
In section \ref{sec:design} we describe our experimental design, in section \ref{sec:results} we discuss the results we obtained and their
impact on computation offloading decisions, and in section \ref{sec:conclusion} we summarize our findings and outline avenues for future work.

\section{Experimental design}
\label{sec:design}
We developed an offloading system testbed for our experiments. This testbed consists of three computers: one models the mobile device that
offloads computation, one is a network appliance that models the network, and one models the server that serves as the offload target. The 
offloading software utilizes Python and matrix arithmetic operations.

The testbed had to meet three important criteria:
\begin{enumerate}
	\item{The client computer has to be significantly slower than the server, so there is a time speedup on remote execution.}
	\item{The bandwidth of the network has to be variable, in order to observe the offloading decision with different communication times.}
	\item{The computational job has to be variable in size, in terms of instructions executed as well as the data moved through the network.}
\end{enumerate}

\subsection{Testbed hardware}
The testbed contains two computers acting as hosts (client and server) and an iBase network appliance to emulate the network. All of these 
computers run Linux, and they are interconnected through 100 Mbps Ethernet:

\begin{itemize}
	\item{Client computer: Intel Pentium 4 CPU at 2.80 GHz, 512 Mb of RAM, running 32-bit Kubuntu 16.04 LTS}
	\item{iBase network appliance: Intel Core 2 Duo E7500 CPU at 2.93 GHz, 8 Gb of RAM, running 64-bit CentOS 7}
	\item{Server computer: Intel i3-2120 CPU at 3.30 GHz, 4 Gb of RAM, running 64-bit Ubuntu 16.04 LTS}
\end{itemize}

The client and the server are connected through the iBase network appliance. The iBase network appliance is a computer with six network 
interface cards and is configured as a software Ethernet switch with software-based traffic control to set a desired bandwidth and latency 
between the client and the server. Manipulating the bandwidth and latency permits us to emulate a multi-hop network with a single device. Open 
vSwitch (OVS)~\cite{web1} provided the Ethernet switching functionality and the \textit{tc} traffic control software provided the bandwidth 
and latency control.

\subsection{Testbed software}
As mentioned above, Open vSwitch (OVS) and \textit{tc} (part of the \textit{iproute2} package) are used on the iBase network appliance to 
create a network emulator. \textit{tc} can artificially delay packets to control latency and can use traffic shaping to control bandwidth. 
We configured \textit{tc} to use the Hierarchical Token Bucket (HTB) queuing discipline~\cite{web2}.

\begin{figure}[htbp]
	\centering
	\includegraphics[scale=0.3]{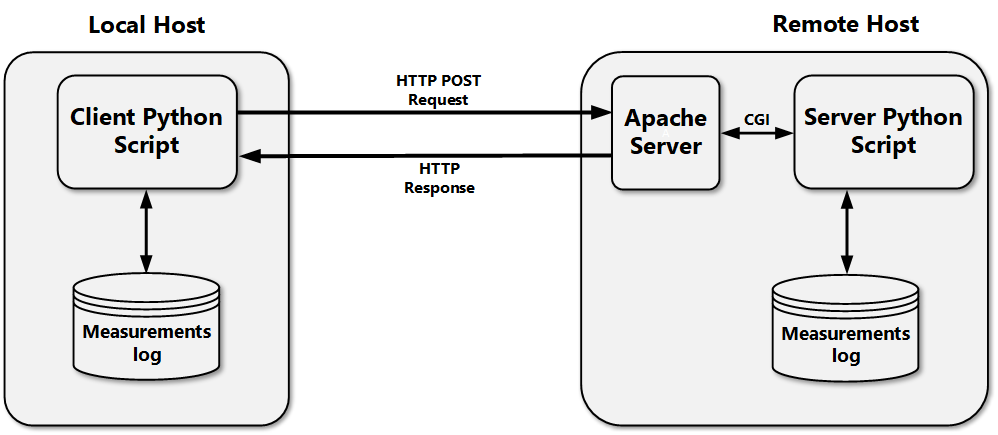}
	\caption[Software architecture of our experimental offloading testbed.]{\scriptsize {Software architecture of our experimental offloading testbed.}}
	\label{fig:swArch}
\end{figure}

The offloading system consisted of a Python script running in the client computer, which can remotely execute intensive computational jobs 
at the server through HTTP requests. Figure \ref{fig:swArch} shows the architecture. The offloading granularity of this system is the function
call. When the client offloads a task, the function identifier and its input parameters are encapsulated in a HTTP request message. The server,
upon receipt of a request, invokes the identified function with the specified input parameters. The server executes Python scripts for the 
function invocation using the Common Gateway Interface (CGI). When the function invocation is complete, the server responds with an HTTP
response message containing the output return value. Figure \ref{fig:swArch} illustrates this system.

\subsection{Testbed instrumentation}
The iPerf3 utility was used to validate the testbed architecture, i.e. to make sure that we actually get the available bandwidth configured 
with the Linux traffic control. This tool measures the bandwidth of a network link by sending data through TCP. 

Python scripts located in the hosts were instrumented with probes to measure the execution time of particular blocks of instructions, based 
on the stages illustrated in Figure \ref{fig:subdivisions}. The probes are implemented with the Python \textit{timeit} module, which 
measures wall-clock time, so the load on the hosts will affect the elapsed time. For this reason, the load on the hosts was reduced while 
running the experiments by disconnecting everything from the Internet and killing unnecessary processes, even the desktop environment. 

\subsection{Set of experiments}
The objective of our experimental study was to understand the impact that the data marshalling technique has on offloading decisions.
Data marshalling is the procedure of moving the input data from the client to the server and the output data from the server to the client.
There are several methods for doing this, as mentioned above we use HTTP's request/response in conjunction with either JSON or raw NumPy 
array data formatting. For JSON, we explore two popular Python implementations: JSON module, and uJSON module.

Our choice of matrix arithmetic for the functions to offload is driven by their high arithmetic intensity which has been shown to make
offloading favorable~\cite{MM2017}. We experimented with several operations: $O$ = $\{Inversion, Natural Logarithm, Multiplication\}$ 
and the \textit{Multiplication} operation provided the most insightful results. We also varied the data size (the dimension of a square 
matrix): $M$ = $\{400, 500, 600, 700, 800, 900, 1000, 1100, 1200\}$ along with the bandwidth of the network (in Mbps): 
$R$ = $\{10, 20, 40, 60, 80, 100\}$.

Each experiment (combination of parameters) were repeated 40 times to permit computation of 95\% confidence intervals.

\subsection{Measures of interest}
The following measures were observed in the experiments:

\textbf{Marshalling-to-Completion time ratio:} Measures the proportion of the marshalling stage compared to the total completion time.

\textbf{Offloading decision vectors:} Binary vectors that represent whether it is convenient to offload (1) or not (0), for each of the 
possible execution cases. Each matrix operation can be represented with a vector whose elements correspond to a particular data rate from 
the set $R$. The sum of the elements for one operation corresponds to the number of convenient offloading cases for all used bandwidths.

\textbf{Alternative method-to-JSON ratio:} This feature shows the proportion between the completion time of either the uJSON or RAW methods 
compared to JSON. 

\textbf{Wrong decision penalties:} It indicates the time penalty when the wrong offloading decision is taken, and it is defined as the 
difference between the remote and local completion times. This metric sets the marshalling methods further apart from each other in case 
their offloading decision vector is the same, as some marshalling methods will incur a smaller time penalty when the offloading decision 
is incorrect. 

\textbf{Cumulative penalty vector:} This metric is computed from the wrong decision penalties, and it is useful for illustration purposes, 
as it is compact to visualize while conveying enough information for comparing each marshalling method. In general, this is a 3-element 
vector for each matrice's size case, where each element represents the summation of the wrong decision penalties along all bandwidths of 
a particular operation using the JSON, uJSON, or RAW marshalling methods: $Penalty = [JSON,uJSON,RAW]$.

\section{Experimental results}
\label{sec:results}
The results presented in this section cover the range of cases where remote execution is definitely not convenient up to the case where 
it is clearly worth to offload the computational job. The remote completion time for some cases is located near to the border of the 
offloading decision, i.e. close to the local completion time, so small changes in the measured times can overturn the benefits of remote 
execution.

\begin{figure}[htbp]
	\centering
	\includegraphics[scale=0.33]{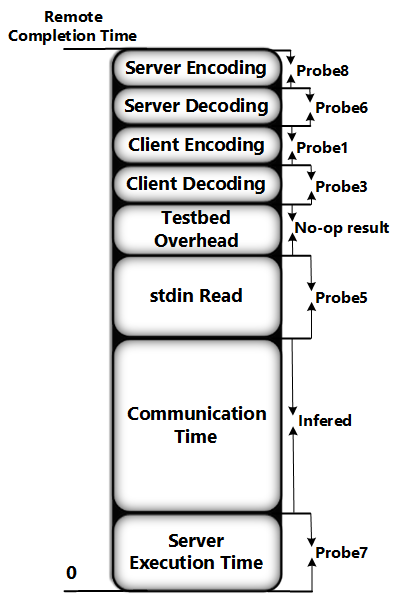}
	\caption[Completion time subdivision for the test cases.]{\scriptsize {The completion time for each test case is subdivided based on the probes instrumented among the offloading code to differentiate one component from another. All stages were measured except for the communication time, which was inferred from the POST request time and the time it take the server to process all of its stages.}}
	\label{fig:subdivisions}
\end{figure}

The remote completion time was subdivided based on the time probes discussed above, and it is composed by the subdivisions illustrated 
in Figure \ref{fig:subdivisions}. The plots presented below follow the same distribution, and they show the bars for each of the bandwidths 
of the set $R$ and operations of the set $O$. Offloading is beneficial when the bar representing the remote completion time under a specific 
bandwidth lays under the dotted line depicting the client execution time.

\subsection{General results on operations with low and medium computational requirements}
The experimental results when $O$ = $\{$Inversion, Natural Logarithm$\}$ can be generalized due to the nature of the operation's 
arithmetic intensity, i.e. the low computation requirements in relation to the data moved through the network. For all cases in this subset 
of operations, the Marshalling-to-Completion time ratio is around the $90\% - 99\%$ and the offloading vectors are equal to zero, 
independently of the bandwidth and the matrice's size. As a result, we showed that it is never convenient to offload these tasks under the 
particular bandwidth conditions.

The Alternative method-to-JSON ratio is rather constant for $O$ = $\{$Inversion, Natural Logarithm$\}$ independently of the matrix size, 
but it changes based on the particular bandwidth. This shows how the alternative methods are faster than JSON, especially RAW data, which 
can be as low as 25\% of the completion time when using JSON.

\subsection{Experimental results when $O$ = $\{$Multiplication$\}$}
In the case of the multiplication operation, we can observe how the computational requirements increase at a faster rate than the data 
moved through the network, so the proposed marshalling performance metrics for computation offloading change over time as the bandwidth
is varied.

\begin{figure*}[!t]
	\centering
	\includegraphics[scale=0.45]{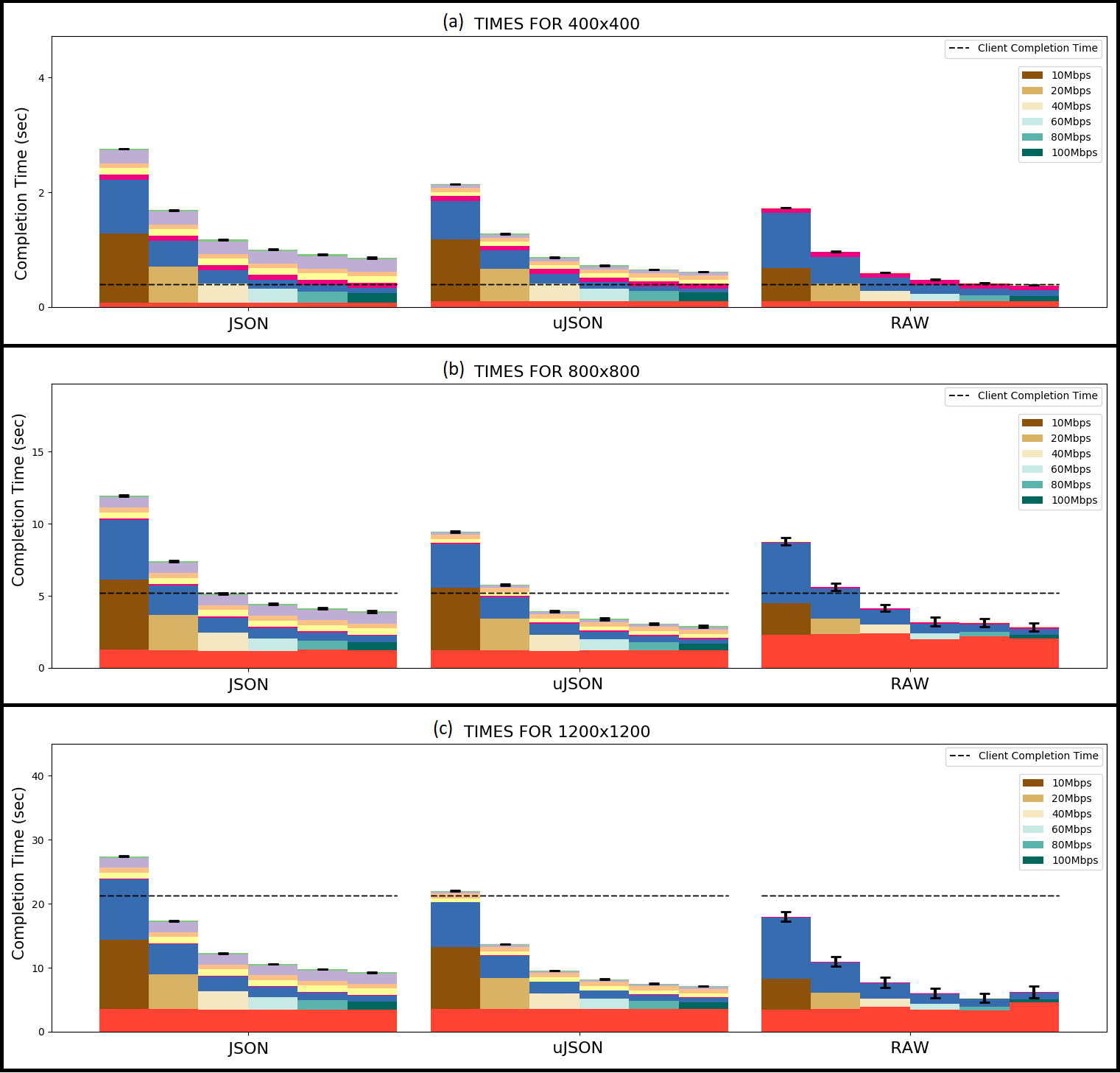}
	\caption[Completion time for various experiments.]{\scriptsize {Completion time for various experiments.}}
	\label{fig:allPlots}
\end{figure*}

In the cases where $S$ = $\{$400$\}$ we have the smallest computing resources requirements due to the matrix dimensions, and it is possible 
to confirm this by looking how the remote completion time is above the local execution time for practically all cases at Fig. 
\ref{fig:allPlots}-(a) , which means that offloading is not worth it. Even then, the remote completion times are closer to the decision 
border when using RAW data. Furthermore, the client execution time is near or above the decoding/encoding times for uJSON, which means that 
by reducing these components the offloading decision may shift. This assumption is indeed confirmed by looking at the RAW method, where the 
offloading time is just below the client execution time because of the marshalling optimization. The confidence intervals are small and 
there is no overlap between the three methods, which proves they are statistically different.

The Marshalling-to-Completion time ratio shows that RAW marshalling is taking up to 72\% of the total completion time, compared to 82\% for 
uJSON and 90\% for JSON. The RAW-to-JSON ratio is around 50\% for all cases at this matrix size, which means that the completion time when 
marshalling with RAW data is twice as fast compared to JSON. As the matrix size increases,  the encoding/decoding times take a considerable 
proportion when using JSON, while these times are lower when using uJSON and barely noticeable when using RAW data. The 
Marshalling-to-Completion time ratios show that the uJSON method starts to fall below 80\% for the highest bandwidths. When $S$ = $\{$600$\}$, 
the execution time increment is more noticeable when using JSON and uJSON, and could be caused by the underlying implementations of these 
methods putting more load on the CPU. The confidence intervals capture this unknown behavior, but the statistical difference between the 
methods is maintained.

When $S$ = $\{$800$\}$, Fig. \ref{fig:allPlots}-(b), the execution time when using RAW data increased with respect to the other methods and 
it shows more variance, a similar behavior as before with uJSON and JSON. This causes the confidence intervals between uJSON and RAW data 
to overlap, but the statistical significance between JSON and RAW data still exists. The marshalling time starts to get significantly lower 
for the alternative methods, going up to 50\% and 25\% of the total completion time for uJSON and RAW data, respectively. As the matrix size 
increase, the abnormal behavior continued, but even then the RAW data method resulted to be the more convenient. The encoding/decoding times 
are still small for RAW compared to the other methods. The Marshalling-to-Completion time ratios are significantly lower for the alternative 
methods at 64\% for JSON and 53\% for uJSON, and especially for RAW data at 24\%.

Finally, in Fig. \ref{fig:allPlots}-(c), when $S$ = $\{$1200$\}$, the remote execution time when using RAW data starts to acquire normal 
values again, even if the confidence intervals are bigger in comparison to the other methods. Moreover, the intervals do not overlap any more 
with the uJSON case, so statistical difference is assured once again for uJSON versus RAW. In this case, uJSON is up to 77\% faster than JSON, 
while the RAW method is up to 67\% faster.

\subsection{Offloading vectors when $O$ = $\{$Multiplication$\}$}
\begin{figure}[htbp]
	\centering
	\includegraphics[scale=0.25]{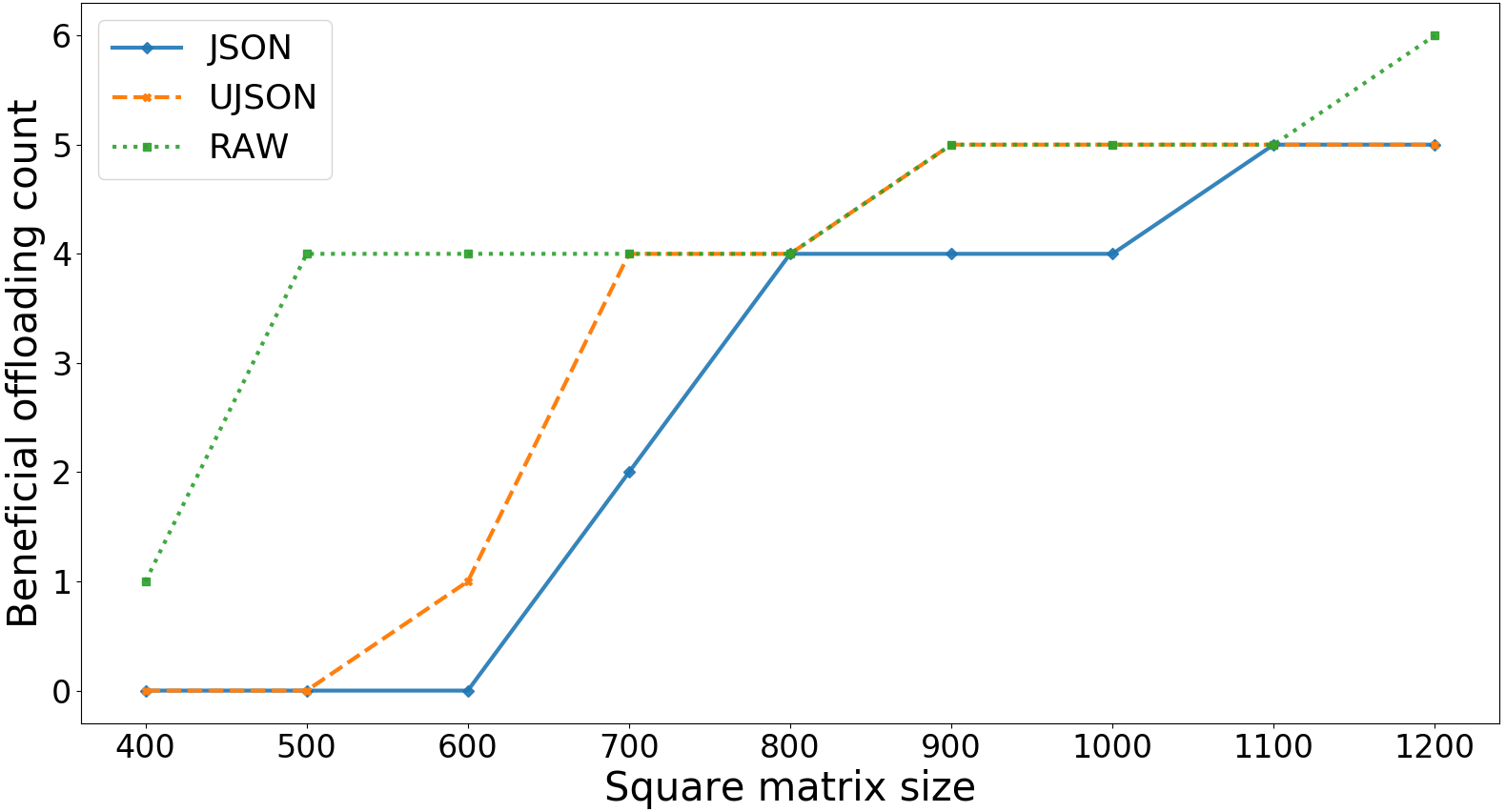}
	\caption[Cumulative penalty vectors when $O$ = $\{Multiplication\}$.]{\scriptsize {Cumulative penalty vectors when $O$ = $\{Multiplication\}$.}}
	\label{fig:offVect}
\end{figure}

In the case of the offloading decision vectors, Fig. \ref{fig:offVect} shows the number of times offloading is convenient for each of the 
three marshalling methods across all the matrix sizes in the set $M$. Even with the smaller matrices, the RAW method for marshalling already 
has one case where offloading is convenient, and the time penalty when making the wrong offloading decision is is 3.34 seconds smaller when 
using RAW data compared to JSON. For the next size, the data method already chooses remote execution for four out of six cases and the 
penalties differences between methods start to be wider apart. 

When $S$ = $\{$600$\}$ the uJSON method has finally a case where offloading is convenient, and the difference between penalties is 10.11 
seconds between RAW and JSON, giving the RAW method a clear advantage over the others. At $S$ = $\{$800$\}$, the offloading decision vector 
is the same for all methods, but it is still possible to differentiate their performance based on the other metrics. The cumulative penalty 
vector shows that RAW data is still the most convenient method in case the offloading decision is incorrectly taken. The penalty time 
difference between RAW and uJSON becomes less than one second, due to the anomaly described above, but even in this case using RAW data is 
more convenient.

When $S$ = $\{$900$\}$, both RAW and uJSON exceeded again the JSON method, and the cumulative penalty vector gives the advantage to RAW 
data for almost one second. For the next matrix size, the offloading vectors remain unchanged but the time penalty between RAW and uJSON 
grows up to 8.34 seconds, giving the RAW method a clear advantage again. Finally, it is always convenient to offload if we are using RAW 
data for the biggest matrix size, and the other two methods are close behind regarding the offloading decision. The difference between the 
time saved by using RAW data continues to grow, giving a difference of 12.94 and 29.38 seconds compared to uJSON and JSON, respectively. 
In conclusion, the offloading decision vectors prove that the RAW method for marshalling is more convenient when performing the offloading 
of tasks with high arithmetic intensity, and the cumulative penalty vector only confirms the benefits of using RAW data over JSON or even 
uJSON.

\section{Conclusions and future work}
\label{sec:conclusion}
We analyzed and dissected the offloading process into its fundamental stages and proposed that the stage normally considered as 
communication time could be further divided to include into data encoding and decoding. These stages in addition to the time the data 
takes to traverse the network, the true communication time, compose the data marshalling time. Our hypothesis was that using different 
methods for data encoding/decoding could impact the offloading decision. An offloading testbed was designed to conduct our experiments 
and sample the necessary time measurements using JSON, uJSON, and RAW techniques for data marshalling.

Several metrics were defined to provide a clear comparison based on the completion time ratios of alternative methods versus JSON, 
the number of cases when offloading is beneficial, and the time penalty if the incorrect offloading decision is taken. In general, we 
observed that remote execution is favored more times when using RAW data compared to the other methods. However, when the matrices, used 
as computational jobs, started to increase in size, the other marshalling method began to approach the performance of RAW data. Even in 
this case, the time penalty of the latter method is smaller when a wrong offloading decision is taken. Overall, our hypothesis was 
true, as we demonstrated how the offloading decision is affected by the marshalling implementation, which will provide a good insight to 
offloading systems design.

The results obtained are general in the sense that they apply for a wide range of offloading systems, so future work along this path is 
the design of a marshalling method optimized for a particular system, for example an ad-hoc network when using fog-computing. We could 
also take a resource scheduling approach to formulate optimization problems when having 
several possible remote execution locations. Finally, another research path is the design of predictors for each factor in the 
offloading decision inequality.

\section*{Acknowledgment}
This work was supported by funding received from Consejo Nacional de Ciencia y Tecnología (CONACyT) through scholarship 439901.

\bibliographystyle{IEEEtran}
\bibliography{compoff}

\end{document}